# Controlling the dynamics of an electric-field-driven droplet on a lubricant-infused micropillar surface

Geng Wang[1,2,3], Junyu Yang[4], Timan Lei[5], Jin Chen[2], Halim Kusumaatmaja[4*], Kai Li[6,1,3*], Kai H. Luo[2*]

1. National Microgravity Laboratory, Institute of Mechanics, Chinese Academy of Sciences, Beijing 100190, China
2. Department of Mechanical Engineering, University College London, Torrington Place, London WC1E 7JE, UK
3. School of Engineering Science, University of Chinese Academy of Sciences, Beijing 100049, China
4. Institute for Multiscale Thermofluids, School of Engineering, The University of Edinburgh, Edinburgh, UK
5. Chair of Building Physics, Department of Mechanical and Process Engineering, ETH Zürich (Swiss Federal Institute of Technology in Zürich), Zürich 8092, Switzerland
6. School of Fundamental Physics and Mathematical Sciences, Hangzhou Institute for Advanced Study, UCAS, Hangzhou, 310024, China

* Corresponding Authors:
halim.kusumaatmaja@ed.ac.uk (H. K.)
likai@imech.ac.cn (K. L.)
k.luo@ucl.ac.uk (K. H. L.)

## Summary

As a non-contact control approach, electric field (EF) can be utilised to drive droplet dynamics on a lubricant-infused surface (LIS), with numerous potential applications ranging from drug manufacturing to 3D printing. However, the resulting droplet dynamics remain poorly understood, especially as there are several possible droplet lubrication states on LIS. Here, we develop a lattice Boltzmann scheme that fully captures the interplay between the interfacial flows and electrohydrodynamics and harness it to investigate EF driven droplets on micropillar LIS. Combining simulations and analytical calculations, we establish quantitative expressions for the drag force and the electric force acting on a moving droplet. We demonstrate that the models can accurately capture droplet dynamics during programmable manipulation, including periodic motion and long-distance transport. Such reliable theoretical models can potentially transform precision control of droplet dynamics by removing the reliance on trial and error tests.

## Introduction

Non-contact manipulation of droplets is critical in a broad range of engineering applications, including digital microfluidics [1,2], chemical reactions [3,4], high-resolution printing [5], and water harvesting [6,7]. In general, droplet manipulation can be classified into two categories. Passive manipulation is achieved by designing geometric, chemical, or wettability-gradient surfaces that drive spontaneous motion towards a lower-energy state [8,9], whereas active manipulation relies on external physical fields such as optical [10], acoustic [11], or electric fields [12,13]. Among these strategies, electric field (EF) based “electric tweezers” (ET) have attracted increasing attention in recent years [14,15]. Compared with other approaches, ET can be applied to most dielectric and leaky-dielectric liquids and requires no additional additives.

The fundamental principle of ET is based on electrohydrodynamics (EHD), where droplet motion results from a balance between electrostatic driving forces and resistive drag. Early ET approaches relied on superhydrophobic surfaces to minimise pinning [12,14,15]. However, control failure can occur when the applied electric field is too strong for a given droplet size [15,16], such that the upward electric force exceeds the surface pinning force, causing the droplet to lift off from the substrate rather than remain in lateral motion. More recently, an ET approach on lubricant-infused surfaces (LIS) has been proposed, thereby removing the reliance on superhydrophobic substrates [17,18]. The presence of the lubricant results in strong suppression of droplet’s contact line pinning and contact angle hysteresis, resulting in highly mobile droplet motion. This strategy significantly broadens the applications of ET, enabling contactless, flexible, and robust droplet manipulation across a wide range of droplet sizes, compositions, and working conditions [17]. However, introducing a lubricant layer substantially increases the complexity of manipulation, as both the electric driving force and the drag force are closely related to the fluid properties, droplet position, and droplet velocity [19,20]. In most practical electric-tweezing applications, droplet manipulation still requires real-time observation of the droplet position to determine electrode actuation [15,17], which is cumbersome and impractical for many applications.

To enable programmable droplet manipulation, theoretical models capable of accurately describing droplet dynamics under different operating conditions are urgently needed [16,21]. Droplet friction on LIS has been widely studied in recent years, and scaling relations between drag force and droplet capillary number have been established [22,23]. Building on this understanding, some experimental studies of electrostatic droplet manipulation have evaluated the electric force using simplified dipole models [17,18,24]. However, existing models generally treat the drag and electric forces separately [25]. With the growing demand for programmable droplet manipulation, such as using time-dependent electrode actuation [17] and different LIS configurations [18], a predictive framework capable of capturing the coupled evolution of electric driving, lubrication state, and interfacial dynamics is required. In summary, two major challenges remain. First, existing expressions for the drag force acting on droplets on LIS are still incomplete, particularly across different lubricant viscosities and surface wettabilities [19]. A primary reason is that the droplet can exist in different lubrication states. Second, a position-dependent description of the electric force acting on a droplet remains unavailable [21].

As a complement to experiments, numerical simulations provide an opportunity to investigate droplet dynamics on LIS over increasingly complex operating conditions. Such numerical simulations are challenging as they must capture complex solid boundaries, three fluid phases (droplet, lubricant, and air), electrohydrodynamics (EHD), and their complex interplay. Previous numerical studies [26,27] have employed molecular dynamics simulations to investigate droplet dynamics on LIS at the nanoscale. For macroscopic droplet transport on LIS, most existing studies have relied on analytical [17] or simplified models [24] to describe droplet friction scaling. However, these models generally provide effective force correlations without directly resolving the transient evolution of different lubrication states. Owing to its mesoscopic nature, the lattice Boltzmann method (LBM) offers inherent advantages in modelling multicomponent multiphase flows [28,29]. Early numerical studies employed ternary LBM to investigate droplet wetting state on LIS [30], followed by subsequent studies of droplet motion control and interfacial dissipation scaling [31,32]. In parallel, LBM frameworks have been developed for two-phase EHD flows based on the leaky-dielectric model [33,34]. Existing studies of EF-controlled droplets on LIS have mainly relied on experiments combined with static or reduced electrostatic calculations [17,24]. A fully coupled numerical framework that simultaneously resolves ternary interfacial flow, charge transport, and the electric field for an EF-driven droplet on an LIS has not previously been reported.

To address these challenges, we develop an EHD ternary multiphase LBM method. This is made possible by bringing together our recently proposed two-phase EHD LBM model for dielectric fluids [35] and multi-component LBM models for droplets on LIS [30,31]. Based on the proposed numerical model, we simulate EF-driven liquid droplets on LIS and then establish a predictive framework that unifies numerical observation and physical mechanism identification with theoretical prediction and programmable manipulation. We first explore the forces acting on the droplet under different operating conditions and use the simulation insights to develop and validate theoretical models for the drag force and the electric force. We show that the resulting models can accurately describe programmable droplet manipulation using pre-designed electrode actuation protocols, including periodic droplet motion and long-distance transport.

# Results

This section establishes a predictive framework for controlling electric-field-driven droplet dynamics on micropillar LIS. Specifically, systematic simulations are first conducted to elucidate how electrode configuration and liquid properties govern the balance between electric driving and interfacial drag. Theoretical models are then formulated for the drag and electric forces and combined into a reduced-order dynamic model that predicts the temporal evolution of the droplet position. This model is subsequently used to design programmable droplet transport strategies.

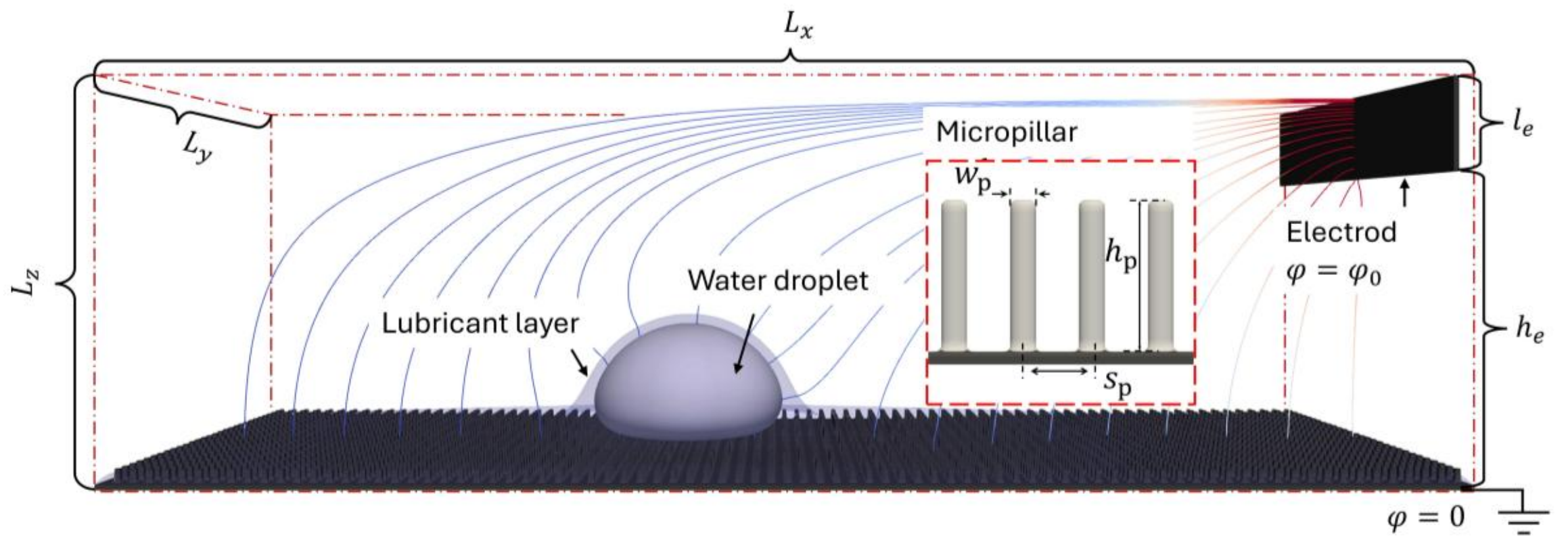

Figure 1. Schematic of computational domain.

Figure 1 shows the physical problem considered in this study: a water droplet is deposited on a silicone-oil-infused micropillar surface and actuated by an electric field generated by an electrode positioned above the substrate near the upper-right corner of the domain. The substrate consists of rectangular micropillars of width $w_{\mathrm{p}} = 42.9\mu\mathrm{m}$, height $h_{\mathrm{p}} = 143.0\ \mu\mathrm{m}$, and edge to edge spacing $s_{\mathrm{p}} = 42.9\ \mu\mathrm{m}$, giving a solid fraction of $\phi = w_p^2/(w_p + s_p)^2 = 0.25$. Unless otherwise specified, the substrate is initially covered by a silicone oil layer with a thickness of $h_{\mathrm{oil}} = h_p$.The water–air, oil–water, and oil–air interfacial tensions are $\gamma_{\mathrm{wa}} = 72.5\ \mathrm{mN/m}$, $\gamma_{\mathrm{ow}} = 19.8\ \mathrm{mN/m}$, and $\gamma_{\mathrm{oa}} = 43.0\ \mathrm{mN/m}$, respectively. The remaining fluid properties are summarised in Table 1. It should be noted that, consistent with the experiments [17], a relatively low-viscosity silicone oil (< 10 cP) was selected as the lubricant for the case of an EF-driven droplet, as a highly viscous lubricant would lead to extremely slow droplet motion when the droplet moves away from the electrode. On a flat solid surface, the intrinsic contact angles of water and oil are denoted by $\theta_{ws}$and $\theta_{os}$, respectively. The electrode configuration is characterised by the electrode-surface distance $h_e$, electrode height $l_e$, and the horizontal separation $S_e$ between the electrode and the droplet.

Table 1 Fluid properties used in the current study

| | Air | Water | Silicone oil |
|---|---|---|---|

| Density | $\rho_g = 1.2$ kg/m$^3$ | $\rho_w = 998.0$ kg/m$^3$ | $\rho_o = 970$ kg/m$^3$ |
|---|---|---|---|
| Viscosity | $\mu_g = 0.0179$ cP | $\mu_w = 1$ cP | $\mu_o = 0.5$~$5$ cP |
| Electrical permittivity | $\varepsilon_g = 8.9 \times 10^{-12}$ F/m | $\varepsilon_w = 80\varepsilon_g$ | $\varepsilon_o = 2.3\varepsilon_g$ |
| Electrical conductivity | $\sigma_g = 1 \times 10^{-50}$ S/m | $\sigma_w = 5 \times 10^{-6}$ S/m | $\sigma_o = 1 \times 10^{-13}$ S/m |

To realise this problem numerically, the simulations are performed in a three-dimensional computational domain of size $L_x \times L_y \times L_z = 7.2 \times 2.6 \times 2.2$ mm. In our simulation, the droplet initial volume is $\text{Vol}_0 = 4\pi R_0^3/3 = 0.33 \sim 4.19\mu L$, corresponding to an effective radius of $R_0 = 0.43 \sim 1$ mm. The micropillar surface is prescribed as zero electric potential, the electrode is maintained at a fixed potential $\varphi_0$, and Neumann boundary conditions are imposed on all other walls. The characteristic electric-field strength is therefore defined as $E_0 = \varphi_0/L_y$. Unless otherwise stated, the spatial resolution is set to $\text{d}x = R_0/35 = 14.3$ μm. A mesh-sensitivity assessment is provided in Supplementary Fig. S1.

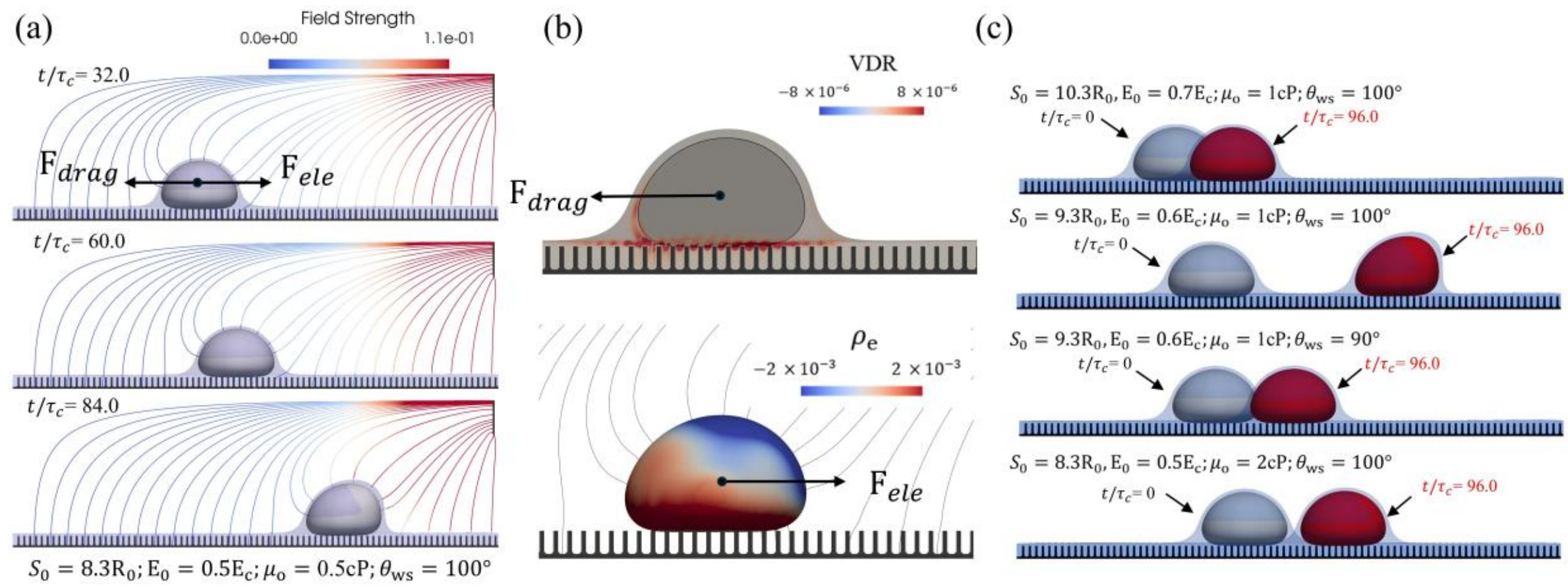

Figure 2. Droplet dynamics under different operating conditions: (a) Droplet migration for the case with $S_0 = 8.3R_0$, $E_0 = 0.5E_c$, $\mu_o = 0.5$ cP, $\theta_{os} = 5°$ and $\theta_{ws} = 100°$. The lines in the figures denote electric field lines. (b) Viscous dissipation (upper panel) and charge density distribution of the moving droplet. (c) Initial (white colour) and terminal (red colour) droplet positions at $t/\tau_c = 96.0$ for different combinations of $S_0$, $E_0$, $\mu_o$ and $\theta_{ws}$.

Figure 2(a) presents typical simulation snapshots of droplet migration. In line with experimental observations [36–38], as the spreading coefficient of oil on water in the presence of air $I_{os(w)} = \gamma_{wa} - \gamma_{ow} - \gamma_{oa} > 0$, the droplet is fully cloaked by the lubricant oil. The droplet starts to migrate directionally (in the electrode direction) under the electric driving force. The droplet migration velocity ($V$) is primarily determined by the competition between the electric force ($F_{ele}$) and drag force ($F_{drag}$) [17]. For droplets moving on LIS, the drag force is dominated by viscous stresses [20]. In the upper panel of Fig. 2(b), we visualise the viscous dissipation of the liquid phases during droplet motion. As $\mu_o < \mu_w$, it can be observed that the dissipation is not only concentrated within the lubricant–droplet meniscus but also occurs inside the droplet (within the black contour), particularly near the micropillar contact region. The inset in the lower panel of Fig. 2(b) shows the charge distribution during droplet migration, where negative charges accumulate on the side facing the electrode due to electrostatic attraction. In addition, inspection of the electric field lines shows that the electric field is only significantly perturbed in the vicinity of the droplet, whereas it remains essentially unchanged far from the droplet throughout its evolution. In this paper, we focus on the effects of the initial droplet–electrode distance ($S_0$), the electric field strength ($E_0$), the lubricant viscosity ($\mu_\mathrm{o}$) and the solid wettability ($\theta_\mathrm{ws}$ and $\theta_\mathrm{os}$) on droplet motion. The first two parameters mainly determine $F_\mathrm{ele}$, whereas the latter two parameters control the magnitude of $F_\mathrm{drag}$. Consequently, the droplet exhibits distinct migration velocities under various combinations of these operating conditions, as illustrated in Fig. 2(c).

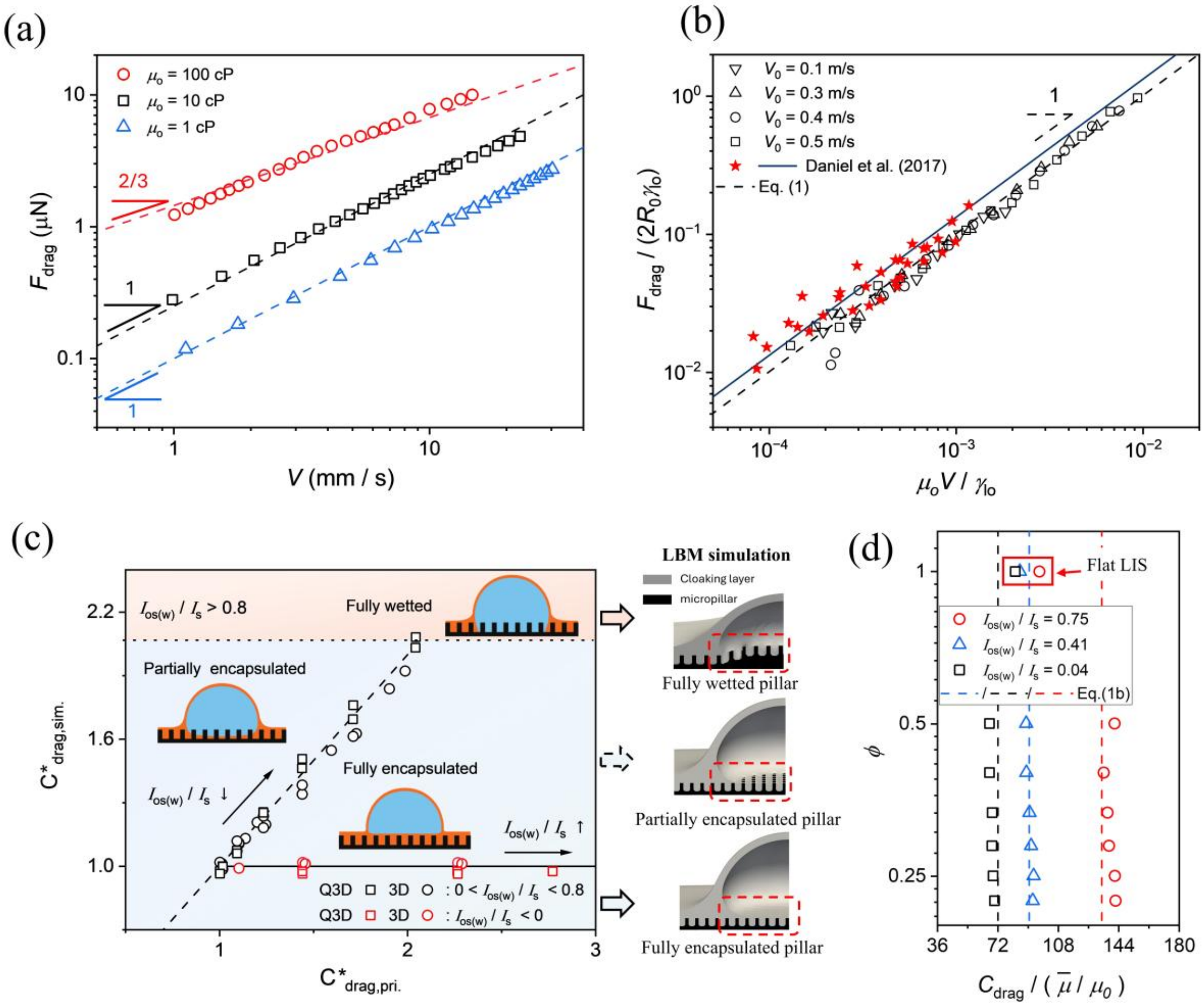

Figure 3. (a) $F_\mathrm{drag}$ as a function of the steady droplet velocity $V$ for lubricant viscosities of $\mu_o = 1$, 10, and 100 cP. Symbols denote the simulation results, and dashed lines indicate the corresponding power-law scalings. (b) The normalized drag force $F_\mathrm{drag}/(2\gamma_{ow}R_0)$ as a function of $\mu_o V/\gamma_{lo}$ during droplet motion, where the red symbols and dashed line denote the experimental data and theoretical equation in Ref. [39]. (c) Comparison between the dimensionless drag coefficient predicted by the proposed theoretical model ($C^*_\mathrm{drag,pri}$) and LBM

simulations ($C^*_{\mathrm{drag,sim}}$). The red shaded region corresponds to $I_{\mathrm{os(w)}}/I_s > 0.8$. In the blue-shaded region, the dashed and solid lines denote $C^*_{\mathrm{drag,sim}} = C^*_{\mathrm{drag,pri}}$ and $C^*_{\mathrm{drag,pri}} = 1$, respectively. In the schematic insets of panel (c), the yellow, blue, and black regions represent the lubricant oil, the droplet, and the micropillars, respectively. The snapshots on the right show representative LBM simulation results for the corresponding lubrication states. (d) Effect of $\phi$ on $C_{\mathrm{drag}}$ for different $I_{\mathrm{os(w)}}/I_{\mathrm{s}}$, symbols represent the simulation results, and dashed lines indicate the theoretical predictions of Eq. (1).

The first issue to be discussed is the drag force ($F_{\mathrm{drag}}$) experienced by a droplet moving on LIS. In recent years, $F_{\mathrm{drag}}$ in the lubricant dissipation dominated regime has been extensively studied [20,25]. In this regime, $F_{\mathrm{drag}}$ increases with the capillary number ($\mathrm{Ca} = \mu_{\mathrm{o}}V/\gamma_{\mathrm{ow}}$) following an approximate $F_{\mathrm{drag}} \propto \mathrm{Ca}^{2/3}$ scaling [22,23]. This behaviour is attributed to viscous dissipation being primarily localised within the Landau-Levich-Derjaguin (LLD) film [39,40], whose thickness scales as $h_{\mathrm{LLD}} \sim R_0\mathrm{Ca}^{2/3}$. The lubricant dissipation dominated regime is typically observed when $h_{\mathrm{LLD}} > h_{\mathrm{p}}$ and $\mu_{\mathrm{o}} \gg \mu_{\mathrm{w}}$. Following previous experiments, we simulated droplets moving on LIS with different lubricant viscosities under driving forces ranging from 0.1 to 10 $\mu$N. As shown in Fig. 3(a), the present model captures the $F_{\mathrm{drag}} \propto V^{2/3}$ (corresponding to $F_{\mathrm{drag}} \propto Ca^{2/3}$) scaling for $\mu_o \gg \mu_w$ ($\mu_o = 100$ cP). This nonlinear scaling indicates that viscous dissipation in the lubricant becomes dominant. For the droplet dissipation dominated regime (i.e., for $h_{\mathrm{LLD}} \ll h_{\mathrm{p}}$ and $\mu_{\mathrm{o}}/\mu_{\mathrm{w}} < 10$), both our simulation results and previous studies [30,39,41] confirmed that $F_{\mathrm{drag}}$ follows a linear dependence on $V$ ($\mu_{\mathrm{o}} = 1$ cP and $\mu_{\mathrm{o}} = 10$ cP in Fig. 3(a), corresponding to $F_{\mathrm{drag}} \propto Ca$).

We also verify this scaling by simulating a droplet deceleration during sliding on a lubricant-infused pillar surface with an initial velocity $V_0$. We test six cases with different $V_0$, $\mathrm{Vol}_0$ and compute $F_{\mathrm{drag}}$ by $F_{\mathrm{drag}} = -mdV/dt$ ( $m = \rho_{\mathrm{w}}\mathrm{Vol}_0$ is droplet mass). The corresponding evolutions of droplet displacement, $S$, and velocity, $V$, can be found in Supplementary Fig. S3. As shown in Fig. 3(b), the calculated drag force $F_{\mathrm{drag}}$ agrees well with both the theoretical prediction $F_{\mathrm{drag}}/(2\gamma_{ow}R_0) \sim \mathrm{Ca}$ and the experimental data reported in Ref. [39] (solid line and red symbols in Fig. 3(b)). The present simulation data in Fig. 3(b) give a slightly smaller best-fit prefactor (dashed line in Fig. 3(b)), compared with the experimental value.

In this paper, we further extend the model of $F_{\mathrm{drag}}$ to a broader range of liquid viscosities and pillar wettabilities, which has not been systematically explored before. For lubricant dissipation dominated cases, $F_{\mathrm{drag}}$ is assumed to scale linearly with $\mu_{\mathrm{o}}$ [25]. However, when dissipation within the droplet is non-negligible, the droplet viscosity must also be considered. We therefore replace $\mu_o$in the original model with an effective viscosity, $(\mu_{\mathrm{o}}\mu_{\mathrm{w}}\mu_{\mathrm{g}})^{1/3}$, defined as the geometric mean of the viscosities of the three fluid phases. To incorporate the effect of pillar wettability, we introduce the interfacial-tension factor $I_{\mathrm{os(w)}}/I_{\mathrm{s}}$. Following Ref. [37], substituting Young's relations $\gamma_{\mathrm{os}} = \gamma_{\mathrm{sg}} - \gamma_{\mathrm{og}}\cos(\theta_{\mathrm{os}})$ and $\gamma_{\mathrm{sw}} = \gamma_{\mathrm{sg}} - \gamma_{\mathrm{wg}}\cos(\theta_{\mathrm{ws}})$ into $\cos\theta_{\mathrm{os(w)}} = (\gamma_{\mathrm{sw}} - \gamma_{\mathrm{so}})/\gamma_{\mathrm{ow}}$ gives the spreading coefficient of oil on the pillar surface in the presence of water, $I_{\mathrm{os(w)}} = \gamma_{\mathrm{og}}\cos(\theta_{\mathrm{os}}) - \gamma_{\mathrm{wg}}\cos(\theta_{\mathrm{ws}}) - \gamma_{\mathrm{ow}}$. We further introduce $I_{\mathrm{s}} = -\gamma_{\mathrm{ow}}$ as a reference quantity associated with the oil–water interfacial tension. Defending $\theta_{\mathrm{ws(o)}}$ as the equilibrium contact angle of water on the solid surface surrounded by oil, Young's relation gives $\gamma_{\mathrm{os}} - \gamma_{sw} = \gamma_{\mathrm{ow}}\cos\theta_{\mathrm{ws(o)}}$, and therefore $I_{\mathrm{os(w)}}/I_{\mathrm{s}} = 1 + \cos\theta_{\mathrm{ws(o)}}$. Based on this interfacial-tension balance, three distinct lubrication states can be identified. Under the present system with $S_{\mathrm{ow(a)}} > 0$ with $I_{\mathrm{os(w)}}/I_{\mathrm{s}} < 0$, the lubricant preferentially wets the solid surface and encapsulates the droplet above the micropillar. For $0 < I_{\mathrm{os(w)}}/I_{\mathrm{s}} < 1$, (i.e.,$180° > \theta_{\mathrm{ws(o)}} > 90°$), the water–oil interface partially penetrates into the pillar gaps, while lubricant is retained in the lower part of the gaps. For $I_{\mathrm{os(w)}}/I_{\mathrm{s}} > 1$, corresponding to $\theta_{\mathrm{ws(o)}} < 90°$, water preferentially wets the solid surface and displaces the lubricant from the pillar surface, resulting in a fully water-wetted pillar state. The three configurations are schematically illustrated in Fig. 3(c). By incorporating effective viscosity and interfacial-tension factor, a new expression for $F_{\mathrm{drag}}$ can be obtained with an empirical fitting coefficient $C_{\mathrm{drag}}$.

$$F^*_{\mathrm{drag}} = \frac{F_{\mathrm{drag}}}{2\gamma_{\mathrm{ow}}R_0} = 3.5\pi C_{\mathrm{drag}}(R_0/l_c)^{0.5}\mathrm{Ca}, \tag{1a}$$

$$C_{\mathrm{drag}} = C_\alpha\left(1 + 1.5\left(\frac{I_{\mathrm{os(w)}}}{I_{\mathrm{s}}}\right)^2\right)\frac{\bar{\mu}}{\mu_{\mathrm{o}}}. \tag{1b}$$

Here, $C_\alpha$ is a fitting constant whose value depends on the simulation configuration (3D or Q3D) and the initial lubricant thickness $h_{oil}$. In the 3D simulations, $C_\alpha \approx 69.8$ for $h_{\mathrm{oil}} = h_p$ and $C_\alpha \approx 63.5$ for $h_{\mathrm{oil}} = 1.2h_p$. In the Q3D simulations, the corresponding values are $C_\alpha = 17.1$ and $C_\alpha = 15.4$, respectively. Thus, the effect of the prescribed lubricant loading is incorporated through $C_\alpha$, whereas progressive lubricant loss and the resulting time-dependent variation in drag during prolonged operation are not considered. The corresponding correlation analysis is provided in the Supplementary Material S2.

To validate the drag model in Eq. (1), we perform simulations at different $\theta_{ws}$, $\theta_{os}$ and $\mu_o$. Considering that the lubricant oil is usually highly wetting on the solid surface for LIS, we restrict our simulations to $\theta_{os} \leq 40°$. Similar to Fig. 3(b), we simulate a droplet deceleration and obtain $S$ from solving deceleration equation of moving droplet $\mathrm{d}^2S/\mathrm{d}t^2 = -F_{\mathrm{drag}}/m$. The comparisons between the theoretical displacement and the numerical results are provided in Supplementary Fig. S4, showing that the proposed model captures the droplet dynamics over the tested ranges of viscosity and surface wettability.

To further assess the generality of the proposed drag model, we perform full 3D and Q3D simulations and compare the simulated drag coefficient, $C_{\mathrm{drag,sim}} = mV_0/(2R_0\bar{\mu}S_{max})$ ($S_{max}$ is the maximum displacement of droplet) with the theoretical prediction $C_{\mathrm{drag,pri}}$ in Eq. (1). A comparison between the normalised $C^*_{\mathrm{drag,sim}}$ and $C^*_{\mathrm{drag,pri}}$ is shown in Fig. 3(c), with both coefficients normalised by $C_{\mathrm{drag}}(I_{\mathrm{os(w)}}/I_{\mathrm{s}} = 0)$. Figure 3(c) reveals three distinct regimes. First, when $I_{\mathrm{os(w)}}/I_{\mathrm{s}} > 0.8$, the lubricant is locally displaced from the pillar surface beneath the droplet, resulting in direct droplet–pillar contact and persistent pinning, as shown in Supplementary Fig. S4. In this pinned regime, the drag relation in Eq. (1) is no longer applicable. This critical value is slightly lower than the theoretical threshold $I_{\mathrm{os(w)}}/I_{\mathrm{s}} > 1$. We suspect that this discrepancy may arise from numerical artefacts associated with the diffuse-interface model used in this study. Second, when $0 < I_{\mathrm{os(w)}}/I_{\mathrm{s}} < 0.8$, $C^*_{\mathrm{drag,sim}}$ agrees well with $C^*_{\mathrm{drag,pri}}$ and increases as $I_{\mathrm{os(w)}}/I_{\mathrm{s}}$ decreases. In this regime, the droplet partially penetrates into the micropillars, as shown in the inset of Fig. 3(c), which is the main reason for the increased drag. Finally, when $I_{\mathrm{os(w)}}/I_{\mathrm{s}} < 0$, the lubricant film beneath the droplet remains continuous and prevents direct contact between the droplet and the micropillars. The droplet therefore slides entirely on the lubricant layer, and the drag coefficient approaches an approximately constant value, close to $C_{\mathrm{drag}}(I_{\mathrm{os(w)}}/I_{\mathrm{s}} = 0)$.

To examine the applicability of the drag model to different micropillar geometries, we extend sliding-droplet simulations in Fig. 3(a) for three different values of $I_{\mathrm{os(w)}}/I_{\mathrm{s}}$ and varying $\phi$. As shown in Fig. 3(d), $C_{\mathrm{drag}}$ exhibits only a weak dependence on the pillar solid fraction when $\phi > 0.2$ and is primarily determined by surface wettability ($I_{\mathrm{os(w)}}/I_{\mathrm{s}}$), similar to the experimental observation in Ref. [42]. The results also agree well with the theoretical prediction of Eq. (1), indicating that the proposed drag model remains applicable over the tested range of micropillar geometries. In contrast, for a flat LIS ($\phi = 1.0$), the values of $C_{\mathrm{drag}}$ obtained at three different values of $I_{\mathrm{os(w)}}/I_{\mathrm{s}}$ collapse to approximately the same value. This limiting case is not captured by Eq. (1), because the micropillar-wetting effect incorporated into $C_{\mathrm{drag}}$ is absent on a flat LIS. Additionally, the lubrication states considered here arise from lubricant displacement within the gaps of a regular micropillar array. For other LIS architectures, such as connected porous networks or disordered microporous substrates, the lubricant retention and transport mechanisms may differ significantly. The applicability of the present drag coefficient–wettability relationship to these structures remains unclear and warrants further systematic investigation.

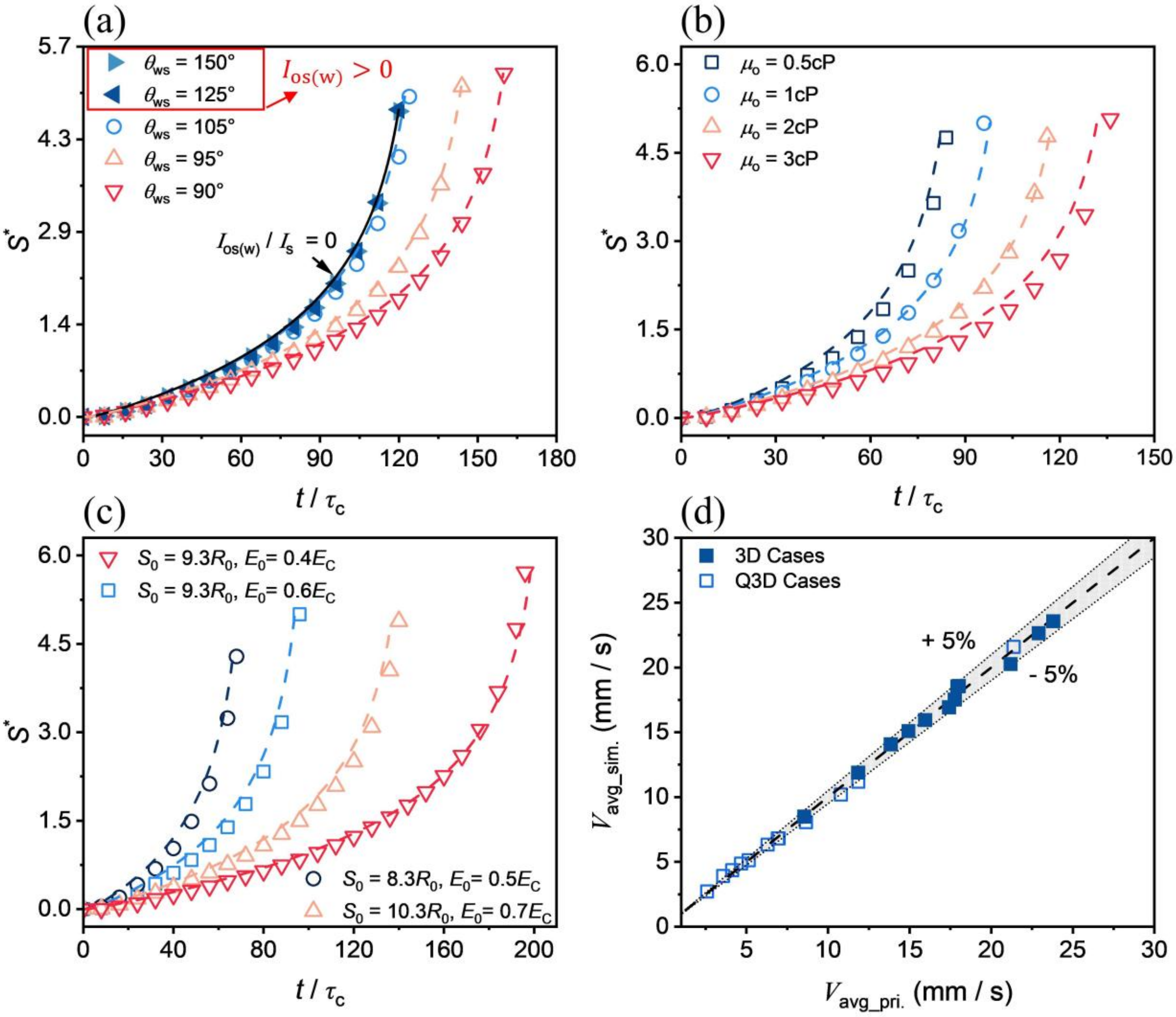

Figure 4. Transient evolution of an EF-driven droplet by varying different operating conditions:(a) droplet–solid apparent contact angle $\theta_{\mathrm{ws}}$; (b) lubricant viscosity $\mu_{\mathrm{o}}$; (c) initial droplet–electrode distance $S_0$ and electric field strength $E_0$, where symbols denote LBM simulation results, lines represent predictions from the theoretical model. Figure (d) is a comparison of the droplet average velocity obtained from LBM simulations and theoretical predictions.

The results in Fig. 3(c) reveal two important mechanisms: (1) The drag force acting on a droplet on LIS can be tuned by adjusting surface wettability, such as $\theta_{\mathrm{os}}$ and $\theta_{\mathrm{ws}}$; (2) When $I_{\mathrm{os(w)}}/I_{\mathrm{s}} < 0$, the drag force becomes independent of surface wettability. This regime can be accessed by varying the droplet, lubricant, and/or solid properties such that contact angles satisfy the inequality $\gamma_{\mathrm{og}}\cos(\theta_{\mathrm{os}}) - \gamma_{\mathrm{wg}}\cos(\theta_{\mathrm{ws}}) - \gamma_{\mathrm{ow}} > 0$ . It should be noted that, consistent with the previous findings [32], we observe that within the present regime of relatively sparse solid fraction ($\phi < 0.25$) and sufficiently tall micropillars ($h_{\mathrm{LLD}} \ll h_p$), variations in pillar height and interpillar spacing do not lead to a noticeable change in $C_{\mathrm{drag}}$. A detailed sensitivity analysis is provided in the Methods section.

We next discuss the electric force acting on the droplet during migration. As the droplet migrates from the far-field region toward the electrode, its local electric capillary number gradually increases, however no appreciable deformation or breakup is observed. When the droplet is sufficiently small compared with the computational domain, the electric force can be approximated using a point-dipole dielectrophoretic (DEP) model [43,44]. Besides, since $\varepsilon_w \gg \varepsilon_o$ and $\sigma_w \gg \sigma_o$, the Coulomb and polarisation forces acting on the lubricant film are negligible compared with those acting on the droplet. For a DC electric field, the time-averaged DEP force on a dielectric sphere can be written as:

$$\begin{cases} F_{\mathrm{ele}} \sim K_e \dfrac{d(|\boldsymbol{E}|_x^2)}{dx}, \\ K_e = 2\pi\varepsilon_{\mathrm{w}} R_0^3 \, \dfrac{\sigma_{\mathrm{w}} - \sigma_o}{\sigma_{\mathrm{w}} - 2\sigma_o}, \end{cases} \tag{2}$$

where $\nabla(|\boldsymbol{E}|_x^2)$ is the gradient of the squared electric field magnitude along the direction of motion. As shown in Fig. 2(a), the droplet perturbs the electric field mainly in the vicinity of the interface. In the far field, the electric potential still satisfies Laplace's equation $\nabla^2\varphi = 0$. Solving it under the boundary conditions detailed in the Supplementary Material yields:

$$F_{\mathrm{ele}} = K_e C_{\mathrm{ele}} e^{-2\alpha_0 S_e}, \tag{3}$$

with:

$$\begin{cases} C_{\mathrm{ele}} = \dfrac{2\alpha_0 \varphi_0^2 \cos^2(\alpha_0 l_e)}{\left(J_{D,n} + \alpha_0 J_{N,n}\right)^2}, \\ \alpha_0 = \dfrac{\pi}{2(h_e + l_e)}, \qquad J_{N,n} = \dfrac{l_e}{2} - \dfrac{\sin(2\alpha_0 l_e)}{4\alpha_0}, \qquad J_{D,n} = \dfrac{h_e}{2} + \dfrac{\sin(2\alpha_0 l_e)}{4\alpha_0}. \end{cases} \tag{4}$$

The analytical solution is derived using a two-dimensional semi-infinite-strip approximation. A detailed derivation of the above analytical solution is provided in the Supplementary Material, together with a comparison with the LBM simulation results presented in Supplementary Fig. S5. It should be noted that both the above derivation and the DEP model assume that the droplet remains weakly deformed and sufficiently small relative to the characteristic length scale over which the background electric field varies. When the initial droplet–electrode distance is $S_0$, combining $F_{\mathrm{drag}}$ with $F_{\mathrm{ele}}$, the equation of motion for an EF-driven droplet on LIS can be written as:

$$m\frac{d^2S}{dt^2} = -C_{\text{drag}} \cdot 2R_0\bar{\mu}\frac{dS}{dt} + \Lambda K_e C_{\text{ele}} e^{-2\alpha_0(S_0-S)}, \tag{5}$$

where $\Lambda$ is a free parameter that balances the relative contributions of the drag force and the electric driving force. We find $\Lambda_{3\text{D}} \approx 0.012$ for 3D cases. With the initial conditions $S(0) = 0$ and $V_0 = 0$, the evolution of droplet displacement can be solved for different operating conditions.

Figures 4(a-c) show the time evolution of the dimensionless droplet displacement ($S^* = S/R_0$) under different $\theta_{\text{ws}}$, $\bar{\mu}$, $S_0$ and $E_0$. Our proposed theoretical model agrees well with the simulation results. Moreover, consistent with the analysis in Fig. 3(b), for cases with $I_{\text{os(w)}}/I_{\text{s}} < 0$, varying the droplet and lubricant wettability have a negligible influence on the droplet migration. To achieve active control of droplet migration, a key requirement is to predict the relationship between migration distance and droplet travel time under various operating conditions [18]. When the droplet is far from the electrode, the viscous relaxation time $\tau_v \sim m/(2C_{\text{drag}}R_0\bar{\mu})$ is much shorter than the timescale over which the electric force varies. Equation (5) therefore admits the approximate analytical form:

$$\frac{dt}{dS} \approx \frac{2C_{\text{drag}}R_0\bar{\mu}}{\Lambda_1 K_e C_{ele} e^{-2\alpha_0(S_0-S)}} + \frac{\alpha_0 m}{C_{\text{drag}}R_0\bar{\mu}}. \tag{6}$$

Integrating Eq. (6) from 0 to $S_t$, the total transport time for a droplet moving can be estimated as:

$$t_{0\sim S_t} = \frac{R_0\bar{\mu}C_{\text{drag}}}{\alpha_0\Lambda_1 K_e C_{ele}}\left(e^{2\alpha_0 S_0} - e^{2\alpha_0(S_0-S_t)}\right) + \left(\frac{\alpha_0 m}{R_0\bar{\mu}C_{\text{drag}}}\right)S_t. \tag{7}$$

Figure 4(d) shows a comparison between the theoretically predicted and simulated average velocities $V_{avg} = S_t/t_{0\sim S_t}$ for droplet travel distance $S_t = 3R_0$. The maximum error of the theoretical prediction (Eq. 7) does not exceed 7% for all cases. We further performed Q3D simulations to assess the generality of the droplet transport model. The theoretical prediction remains valid for the Q3D cases; the only difference is that the fitting parameter $\Lambda_{1,\text{Q3D}} = \Lambda_{1,3\text{D}}/2$.

To evaluate the predictive capability of the theoretical model, we applied Eq. (7) to the experimental conditions reported in Ref. [17]. For a 12 μL water droplet with $S_0$ of 9.45 mm, an applied voltage of 5.5 kV, and a lubricant viscosity of 10 cP. In this simulation, we set $\theta_{\text{ws}} = 110°$ and $\theta_{\text{os}} = 5°$, corresponding to $I_{\text{os(w)}}/I_{\text{s}} \approx 0$. The simulated droplets exhibit a measured apparent contact angle of approximately 95°, which is consistent with the experiment. The predicted droplet travel time ($S_t = S_0$) by Eq. (7) is approximately 9.41s, with a deviation of only approximately 3% from the experimental measurement. Figure 5(a) compares the temporal evolution of the normalized droplet position obtained from the experiment, LBM simulation, and theoretical model. The displacement and time are normalized by $S_0$ and $t_{0\sim S_0}$, respectively. Good agreement is observed among the three results, although our theoretical model and LBM simulation results slightly underestimate the displacement during $0.5 < t/t_{0\sim S_0} < 0.8$. This difference may be attributed to differences in the LIS structure, particularly the experimentally unreported exact value of $\theta_{\mathrm{ws}}$ and $\theta_{\mathrm{os}}$. The simulated droplet morphology and its position relative to the electrode at different normalized times also agree qualitatively with the experimental observations. These comparisons confirmed that the proposed theoretical model can accurately predict the time required for the droplet to reach a prescribed position.

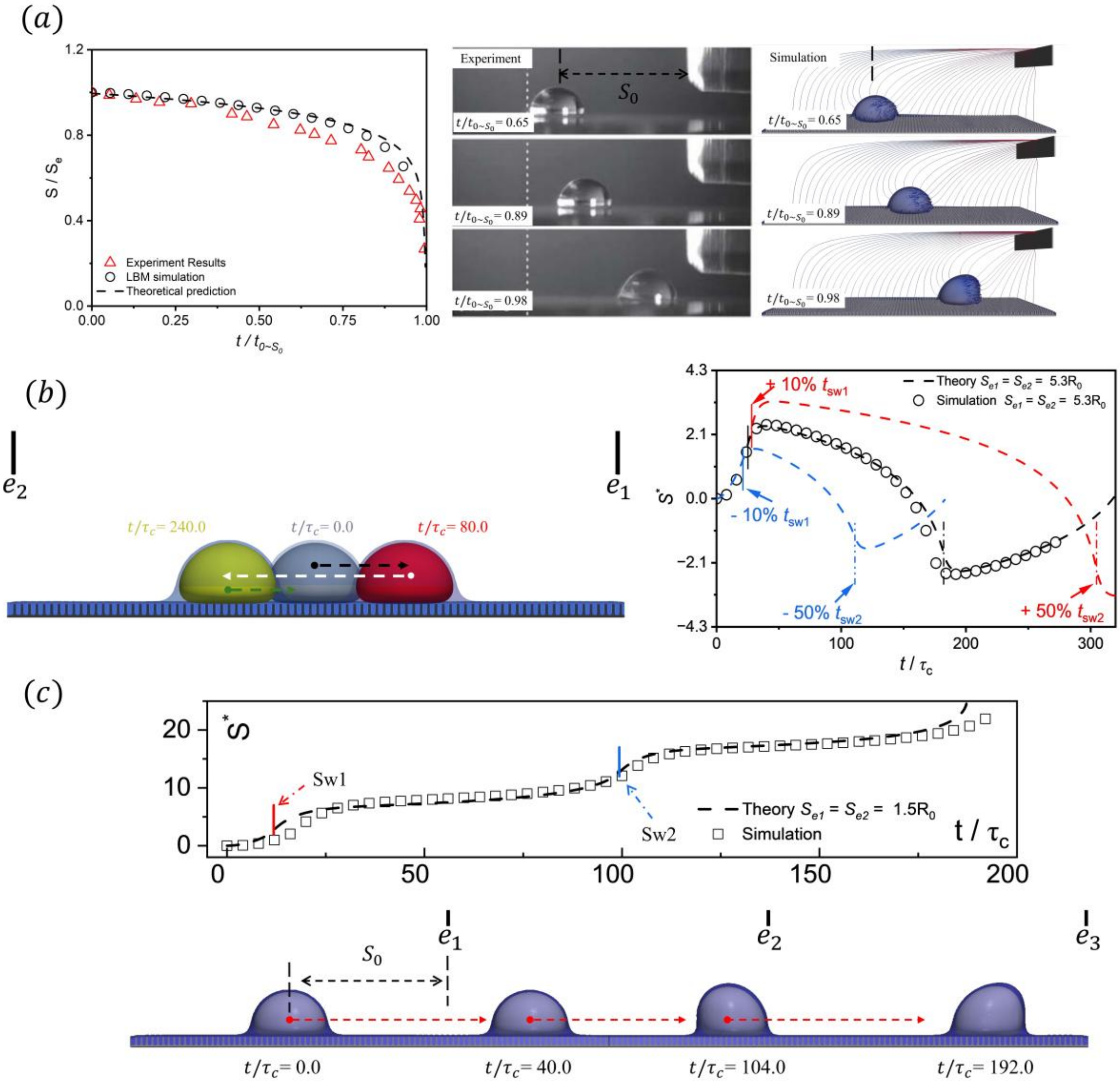

Figure 5. Simulation of directional droplet motion driven by an electric field. (a) Comparison of the experimental, LBM, and theoretical results for droplet migration toward the electrode. The left figure shows temporal evolution of the normalized droplet displacement, and the right panel shows experimental and simulated droplet morphologies and positions relative to the electrode at selected normalized times. (b) Left: qualitative snapshots

of periodic droplet motion, with different colours corresponding to different time instants. Right: quantitative comparison between the simulated droplet displacement and the theoretical prediction. (c) Top: quantitative comparison between the simulated droplet displacement and the theoretical prediction for long-distance droplet transport. The bottom panel shows representative snapshots.

Building on our theoretical model, we enable active control by pre-designing electrode switching protocols for different transport modes. In this work, we propose two electrode-control strategies to realise periodic and long-distance droplet transport. First, we place a droplet in the middle of two electrodes and control its motion by setting the switching times of electrodes $e_1$ and $e_2$. We initially activate $e_1$ to drive the droplet. When the droplet approaches electrode $e_1$ such that the droplet–electrode distance $S_{e1}$ decreases to $4.5R_0$, we switch off $e_1$ and activate $e_2$. When the droplet subsequently moves back and the droplet–electrode distance to $e_2$, denoted by $S_{e2}$, decreases to $4.5R_0$, we switch off $e_2$ and reactivate $e_1$. The switching times are obtained by solving Eqs. (5) and (7). Representative snapshots of the bidirectional droplet motion are shown in Fig. 5(b), together with the corresponding quantitative evaluation of droplet displacement. In addition, an animation of the bidirectional droplet is provided in Supplementary Movie 1. Our LBM simulations confirm that periodic droplet motion can be achieved through electrode switching, and the simulated displacement agrees closely with the theoretical prediction. It is worth noting that a small deviation in the switching time can significantly alter the mode of droplet migration. As an illustration, we plot the theoretical displacement curves obtained when the first switching time ($t_{\mathrm{sw1}}$) is advanced by 10% or delayed by 10%. The corresponding time for the droplet to reach the second switching point ($t_{\mathrm{sw2}}$) differs by approximately 50%. This sensitivity highlights that an accurate theoretical model is essential for achieving reliable active control of droplet transport.

Finally, similar to the experiment in Ref. [15] , we simulate long-distance droplet transport driven by multiple electrodes. We consider three electrodes separated by a fixed spacing $S_e$, with the initial droplet–electrode distance set to $S_0 = 5R_0$. The electrodes are switched when the droplet approaches the active electrode to $S_{e1} = S_{e2} = 1.5R_0$. Considering the substantial computational cost of this simulation, we adopted a thicker initial lubricant layer ($h_{\mathrm{oil}} = 1.2h_p$) in this case to reduce $F_{\mathrm{drag}}$ (as discussed in Supplementary Material), thereby shortening the simulation time. The animation of the long-distance droplet transport is shown in Supplementary Movie 2. Similar to the case in Fig. 5(b), the switching times are obtained by solving Eqs. (5) and (7). Figure 5(c) presents the quantitative evolution of the droplet displacement together with representative snapshots. The results show that the theoretical model accurately predicts the droplet displacement for this long-distance droplet transport problem. A small deviation can be observed between the theoretical model and the simulation results at the initial stage ($t/\tau_{\rm c} < 25$) and final stage ($t/\tau_{\rm c} > 175$). This discrepancy arises from a limitation of the point-dipole dielectrophoretic model used to evaluate the electric force, which is derived under the assumption of an infinite domain. In the present case, the electrode is initially located close to the droplet and the electrode remains active at the final stage, so finite size effects of the droplet are no longer negligible. As a result, the point-dipole approximation becomes less accurate, leading to the observed deviation.

# Conclusion

In this work, we conducted a systematic investigation into the dynamics of EF-driven droplets on LIS and revealed the competition of forces governing droplet motion under different operating conditions. Our results show that the lubrication state and the drag force of a droplet on LIS can be tuned by modifying the surface wettability of the solid substrate. The interfacial tension factor $I_{\rm os(w)}/I_{\rm s}$ can be used to characterise the droplet lubrication state on LIS. Then, we extend existing LIS drag descriptions to a broader range of lubricant viscosities and pillar wettabilities, which have not been systematically explored previously. In addition, through analytical derivation, we proposed a position-dependent electric force model that accounts for applied electric field strength and droplet position. Their combination establishes a reduced-order predictive framework through which the transient droplet displacement and transport time can be rapidly determined from the prescribed electrode configuration, fluid properties and surface wettability. The predicted transport time can then be used to determine the required electrode-switching times, providing a physics-based tool for designing programmable droplet-transport strategies prior to experimental implementation. Based on the proposed theoretical framework, we designed programmable droplet manipulation by pre-setting the electrode-switching times and electrode arrangement, enabling periodic droplet motion and long-distance droplet transport.

One important outcome of this study is the establishment of the quantitative relationship between drag force and surface wettability, providing new insight for the precise control of droplet dynamics on LIS. Based on this theoretical framework, droplet motion can be predicted within the calibrated parameter range once the physical properties of the droplet and lubricant, the wettability of the micropillar surface, and the electrode position and field strength are specified. This work reduces the reliance of EF-driven droplet manipulation on LIS on continuous monitoring of droplet position to adjust the electrode configuration and opens up new avenues to droplet manipulation in applications such as 3D printing, printed electronics, microfluidics, and drug delivery. Recent studies of interfacial force regulated solid like liquid sliders and interfacial droplet-based triboelectric nanogenerators further highlight the importance of interfacial-force regulation, wetting confinement and electrode arrangement for solid–liquid electrical systems [45,46]. As a future task, the proposed manipulation strategy could be experimentally implemented using individually addressable positioned above or beside a grounded micropillar-array LIS, with programmable high-voltage switches applying the predicted switching sequence. Device-specific calibration of the electric-force and drag coefficients would be required, together with control of the droplet volume, conductivity and initial position, lubricant loading and surface wettability, and switching accuracy. Since the present demonstrations involve no more than three switching events, cumulative surface-charge accumulation, lubricant redistribution and interfacial degradation are negligible. Such effects during long-term repeated actuation remain to be investigated.

# Method

### Governing equations for the ternary multiphase EHD model

Electric-field-driven droplets on LIS represent a typical ternary multiphase EHD problem. Fully simulating EF-driven droplet motion on an LIS presents three major challenges. First, a ternary multiphase model that accurately represents the interfacial tensions, wetting boundary conditions, and moving contact lines is required. Second, the leaky dielectric model (LDM) [47] requires charge transport and the resulting Maxwell stresses to be coupled with the evolving interfaces and flow field. Third, the electric-potential equation with spatially varying electrical properties requires inner iterations at each hydrodynamic time step. The present framework addresses these challenges by coupling a mesoscopic ternary LBM with a charge-transport-based LDM. Before constructing the numerical model, several assumptions are specified. First, the three fluid phases are assumed to be incompressible and immiscible. Second, compressibility effects are neglected, and each phase is treated as a Newtonian fluid with constant properties. Third, surface charges are treated as volumetric charges distributed within the interfacial diffusion layer, and no free charges are assumed to exist in the bulk fluids.

In the present model, a phase-field approach is adopted to simulate the ternary multiphase flow, and an improved conservative Allen–Cahn (AC) equation is employed for interface tracking [48]:

$$\frac{\partial \Phi_i}{\partial t} + \nabla \cdot (\Phi_i \boldsymbol{u}) = M_i \nabla \cdot \left( \nabla \Phi_i - \frac{4\mathbf{n}_i}{W_{int}} |\Phi_i(1-\Phi_i)| + \frac{\Phi_i^2}{\sum_{j=1}^{3} \Phi_j^2} \sum_{j=1}^{3} \frac{4\mathbf{n}_j}{W_{int}} |\Phi_j(1-\Phi_j)| \right), \quad (8)$$

where $\mathbf{u} = [u_x, u_y, u_z]$is the fluid velocity. $\Phi_i (i = 1,2,3)$ denote the phase indicators of the ternary components and satisfy the constraint $\Phi_1 + \Phi_2 + \Phi_3 = 1$. The phase interfaces are identified by $\Phi_i = 0.5$. $\mathbf{n}_i = \nabla\Phi_i/|\nabla\Phi_i|$ is the unit normal vector to the interface of component $i$. $M_i$ represents the mobility, and $W_{\mathrm{int}}$ denotes the interface thickness. Following Ref. [49], to ensure numerical stability and accurate interface resolution at large density ratios, we set $M_1 = M_2 = M_3 = 0.0025$ and $W_{\mathrm{int}} = 5$ in lattice units.

In the phase field model, the conservative AC equation is coupled with incompressible NS equations for the fluid flow, where [50]:

$$\nabla \cdot \boldsymbol{u} = 0,$$
$$\frac{\partial(\rho\boldsymbol{u})}{\partial t} + \nabla \cdot (\rho\boldsymbol{u}\boldsymbol{u}) = -\nabla P + \nabla \cdot \left(\rho\nu(\nabla\boldsymbol{u} + \nabla\boldsymbol{u}^T) + \rho\left(\nu_b - \frac{2}{3}\nu\right)(\nabla \cdot \boldsymbol{u})\mathbf{I}\right) + \boldsymbol{F_s} + \boldsymbol{F_e}. \tag{9}$$

$\rho$ denotes the fluid density, $\nu$ is the kinematic viscosity, $\nu_b$ represents the non-hydrodynamic bulk viscosity, and $P$ is the pressure. For the ternary phase-field model, the surface tension force is given by $\boldsymbol{F_s} = \mu_{\Phi_i}\nabla\Phi_i$, and the ternary chemical potential is defined as [51]:

$$\mu_{\Phi_i} = \frac{24\Upsilon_i}{W_{int}}\Phi_i(\Phi_i - 1)(\Phi_i - 0.5) - \frac{24\Upsilon_T}{W_{int}}\Phi_1\Phi_2\Phi_3 - \frac{3}{4}\Upsilon_i W_{int}\nabla^2\Phi_i, \tag{10}$$

$\Upsilon_T$ and $\Upsilon_i$ are parameters related to the surface tension:

$$\begin{gathered} 3\Upsilon_T^{-1} = \Upsilon_1^{-1} + \Upsilon_2^{-1} + \Upsilon_3^{-1}, \\ \Upsilon_1 = \gamma_{12} + \gamma_{13} - \gamma_{23}, \\ \Upsilon_2 = \gamma_{12} + \gamma_{23} - \gamma_{13}, \\ \Upsilon_3 = \gamma_{13} + \gamma_{23} - \gamma_{12}. \end{gathered} \tag{11}$$

The above ternary multiphase model is coupled with the LDM model for EHD problem. In LDM, the electric force ($\boldsymbol{F_e}$) can be expressed as [47]:

$$\boldsymbol{F_e} = \rho_e\mathbf{E} - \frac{1}{2}\mathbf{E}^2\nabla\varepsilon, \tag{12}$$

where $\mathbf{E} = -\nabla\varphi$ is electric field strength. $\varphi$ stands for the electric potential, $\varepsilon$ is dielectric permittivity, $\rho_e$ is electrical charge density and fluid density. The governing equation of the electric field follows Gauss's law:

$$\nabla \cdot (\varepsilon\nabla\varphi) = -\rho_e. \tag{13}$$

The charge transport equation is used to describe the evolution of charge density [52,53], where:

$$\frac{\partial\rho_e}{\partial t} + \nabla \cdot (\rho_e\boldsymbol{u}) - \nabla \cdot (\sigma\nabla\psi) - \alpha\nabla^2\rho_e = 0, \tag{14}$$

The dimensionless charge diffusion number is $\alpha^* = \alpha\mu_w/(\varepsilon_w \mathrm{R}_0^2\mathrm{E}_0^2) = 10^{-4}$ in current simulation, and $\sigma$ is the electrical conductivity of the fluid. The introduced leaky dielectric model (LDM) has been extensively adopted and validated in our recent studies. It has demonstrated the capability to accurately reproduce a broad range of challenging EHD problems, including Rayleigh fission, electrospray, droplet equatorial streaming, and EF enhanced pool boiling.

**Unified Lattice Boltzmann Model for the ternary multiphase EHD model**

In this work, the Unified Lattice Boltzmann Model (ULBM) proposed by Luo et al. [54] is employed to solve the governing equations presented in the preceding sections. ULBM is a unified framework that incorporates different collision models, multiphase flow models, and force schemes. To recover the target AC equation, a D3Q17 multi-relaxation time (MRT) ULBM is constructed, the collision operator of ULBM in raw moment space can be written as:

$$\mathbf{m}_{\Phi_i}^{*} = \left(\mathbf{I} - \mathbf{S}_{\Phi_i}\right)\mathbf{m}_{\Phi_i} + \mathbf{S}_{\Phi_i}\mathbf{m}_{\Phi_i}^{\mathbf{eq}} + \Delta t\left(\mathbf{I} - \frac{\mathbf{S}_{\Phi_i}}{2}\right)\mathbf{M}C_{\Phi_i}, \tag{15}$$

where $\mathbf{m}^{*}$, $\mathbf{m}^{\boldsymbol{eq}}$ are post-collision distribution moment set and equilibrium moment set, respectively. $\mathbf{I}$ stands for the unit matrix and $\mathbf{S}$ is the relaxation matrix. The subscript $\Phi_i$ represents the phase indicator, $\boldsymbol{e}_n$ and $\Delta t = 1$ are the discrete velocities and the time step, respectively. In this study, the transformation matrix $\mathbf{M}$ is chosen as the simplified non-orthogonal moment set which was originally proposed by Fei et al. [55,56]. The equilibrium distribution function for component $i$ can be as:

$$f_{\Phi_i,n}^{eq} = \Phi_i \omega(|\boldsymbol{e}_n|^2)\left[1 + \frac{\boldsymbol{e}_n . \boldsymbol{u}}{c_s^2}\right], \tag{16}$$

where the weighs are $\omega(0) = 1/3$, $\omega(1) = 1/18$, and $\omega(2) = 1/36$ for D3Q19 model, and $c_s = 1/\sqrt{3}$ stands for the lattice sound speed. Multiplying the transformation matrix $\mathbf{M}$, the raw moment space discrete equilibrium moment is $\mathbf{m}_{\Phi_i}^{\mathbf{eq}} = \mathbf{M}f_{\Phi_i,n}^{eq}$, the forcing term can be written as [57]:

$$C_{\Phi_i,n} = \frac{\omega(|\boldsymbol{e}_n|^2)}{c_s^2}\boldsymbol{e}_n\left(\boldsymbol{F}_{\Phi_i,n} + \frac{\partial(\Phi_i \boldsymbol{u})}{\partial t}\right), \tag{17}$$

$\boldsymbol{F}_{\Phi_i,n}$ is given by:

$$\boldsymbol{F}_{\Phi_i,n} = \left[F_{\Phi_i,x}, F_{\Phi_i,y}, F_{\Phi_i,z}\right] = \frac{4\mathbf{n}_i}{W_{int}}|\Phi_i(1-\Phi_i)| - \frac{\Phi_i^2}{\sum_{j=1}^{3}\Phi_j^2}\sum_{j=1}^{3}\frac{4\mathbf{n}_j}{W_{int}}\left|\Phi_j\left(1-\Phi_j\right)\right|, \tag{18}$$

and the phase indicators can be calculated by:

$$\Phi_i = \sum_n f_{\Phi_i,n}, \tag{19}$$

The distribution function for $\Phi_i$ ($f_{\Phi_i,n}$) is reconstructed by $f_{\Phi_i,n} = \mathbf{M}^{-1}\mathbf{m}_{\Phi_i}^{*}$ after the collision in the raw moment space. To model the wetting behaviours involved in the simulation, we employed an energy-based contact angle model, in which the relationship between the solid phase indicator ($\Phi_{i,s}$) and the apparent contact angle ($\theta$) can be expressed as [58]:

$$\boldsymbol{n}\cdot\nabla\Phi_{1,s} = \frac{4}{W_{int}}\left(-\cos(\theta_{13})\Phi_{1,s}\Phi_{3,s} - \cos(\theta_{12})\Phi_{1,s}\Phi_{2,s}\right),$$
$$\boldsymbol{n}\cdot\nabla\Phi_{2,s} = \frac{4}{W_{int}}\left(\cos(\theta_{12})\Phi_{1,s}\Phi_{2,s} - \cos(\theta_{23})\Phi_{2,s}\Phi_{3,s}\right), \quad (20)$$
$$\boldsymbol{n}\cdot\nabla\Phi_{3,s} = \frac{4}{W_{int}}\left(\cos(\theta_{23})\Phi_{2,s}\Phi_{3,s} + \cos(\theta_{13})\Phi_{1,s}\Phi_{3,s}\right).$$

$\theta_{ij}$ stands for the contact angle of the interface shaped by fluid $i$ and $j$ with the solid surface.

The incompressible Navier–Stokes equations, the charge conservation equation, and the Poisson equation are solved using a methodology similar to that adopted in our recent studies [35,59]. For example, two different distribution functions are introduced:

$$\mathbf{m}_g^* = \left(\mathbf{I} - \mathbf{S}_g\right)\mathbf{m}_g + \mathbf{S}_g\mathbf{m}_g^{eq} + \Delta t\left(\mathbf{I} - \frac{\mathbf{S}_g}{2}\right)\mathbf{R}_g,$$
$$\mathbf{m}_{\rho_e}^* = \left(\mathbf{I} - \mathbf{S}_{\rho_e}\right)\mathbf{m}_{\rho_e} + \mathbf{S}_{\rho_e}\mathbf{m}_{\rho_e}^{\mathrm{eq}} + \Delta t\left(\mathbf{I} - \frac{\mathbf{S}_{\rho_e}}{2}\right)\boldsymbol{R}_{\rho_e} + \Delta t\boldsymbol{C}_{\rho_e} + 0.5\Delta t^2\partial_t\left(\boldsymbol{C}_{\rho_e}\right), \quad (20)$$

where $\mathbf{m}_g = \mathbf{M}g_n$ is associated with the incompressible Navier–Stokes equations, and $\mathbf{m}_{\rho_e} = \mathbf{M}f_{\rho_e}$ corresponds to the charge conservation equation. After the collision step in raw moment space, the distribution functions are reconstructed as $g_i = \mathbf{M}^{-1}\mathbf{m}_g^*$ and $f_{\rho_e} = \mathbf{M}^{-1}\mathbf{m}_{\rho_e}^*$. For the Poisson equation, an inner iteration is required at each time step to reach the steady state. To reduce the computational cost, a D3Q7 single-relaxation-time collision operator is adopted to solve the Poisson equation:

$$f_{\psi,i}^* = f_{\psi,i} - \frac{1}{\tau_\psi}\left(f_{\psi,i} - f_{\psi,i}^{eq}\right) + \Delta t' C_{\psi,i} + 0.5\Delta t'^2\partial_t\left(C_{\psi,i}\right), \quad (21)$$

where $\Delta t' = 1$ is the inner iteration time step. In the above equations, the discrete equilibrium moment sets and the discrete forcing terms, including $\mathbf{m}_{\rho_e}^{\mathrm{eq}}$, $\mathbf{m}_g^{\mathrm{eq}}$, $\mathbf{R}_g$, $\mathbf{R}_{\rho_e}$, $\mathbf{C}_{\rho_e}$, $f_{\psi,i}^{\mathrm{eq}}$ and $\mathbf{C}_{\psi,i}$, are given explicitly in our previous work studies [35,59] According to the Chapman–Enskog analysis in our recent studies [35,60], it has been demonstrated that the above LBM model can accurately reproduce the incompressible NS equations, the charge conservation equation, and the Poisson equation. The detailed derivations are not repeated here for brevity. The simulation code is parallelized using MPI, and a typical case that models EF-driven droplet dynamics requires approximately 2500 CPU cores running for more than 10 h.

**Model validations**

In this work, the proposed model is comprehensively validated by comparing our simulations with available experimental and simulation studies. It should be noted that the EHD multiphase model employed here has been extensively validated in our recent studies [35,59]. Therefore, we focus on validating the additional ternary multiphase model in this section. First, we reproduce the experiments of Planchette et al. [61], in which a glycerol-solution droplet and a silicone-oil droplet with the same initial diameter $D_0$ undergo head-on collisions with different relative velocities $V_0$ and off-centre distances $X$.

We simulate two cases with different impact parameters: (a) $D_0 = 230$ μm, $V_0 = 2.30$ m/s, $X = X_{off}/D_0 = 0.03$ and (b) $D_0 = 190$ μm, $V_0 = 2.75$ m/s, $X = X_{off}/D_0 = 0.49$, $X_{off}$ is the transverse offset between the two droplet-centre trajectories. As shown in Fig. 6, the numerical results agree well with the experimental observations [62]. In particular, for the higher impact velocity case, the simulations successfully capture the "cross-separa.tion" mechanism reported experimentally, where only a pure droplet of the encapsulating liquid remains after collision, together with the formation of satellite droplets.

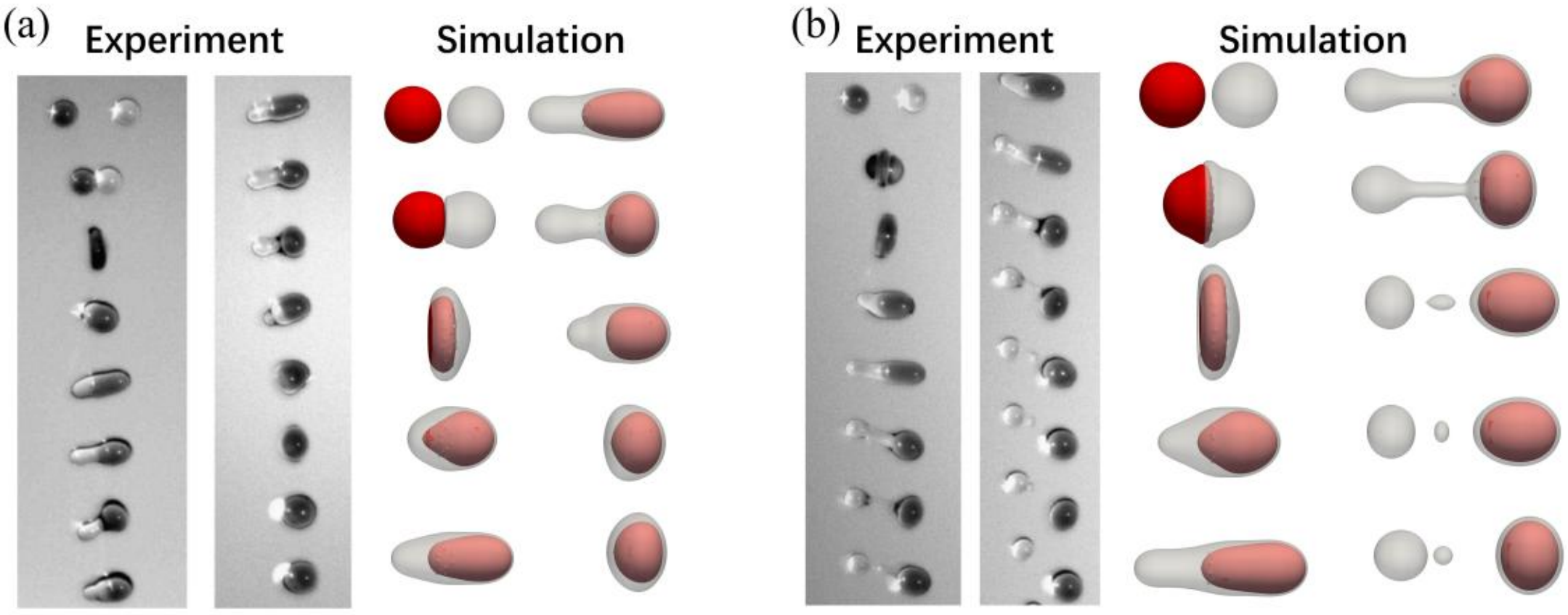

Figure 6. A qualitative comparison between experimental and numerical results for head-on collisions of a glycerol-solution droplet and a silicone-oil droplet.

Next, we validate the translational mobility of droplets on LIS, following previous experimental [41] and numerical studies [30]. We reproduce droplet sliding on LIS driven by a fixed body force ($F_{\mathrm{x}} = mg_x$), using the similar configuration ($\phi = 0.25$, Bond number $\mathrm{Bo} = R_0^2 g_x/\gamma_{\mathrm{wg}} = 0.112$ and apparent angle $\theta_{\mathrm{app}} = 90°$) to experiment in Ref. [41]. In this case, the lubricant viscosity is set to $\mu_{\mathrm{o}} = 10\mu_{\mathrm{ref}}$, where $\mu_{\mathrm{ref}} = 1\mathrm{cP}$ is the reference viscosity. The droplet viscosity is varied over the range $\mu_{\mathrm{w}} = 1 \sim 500$ cP. Similar to Ref. [30], we scale the droplet terminal velocity $V_{\mathrm{t}}$ by $V_{\mathrm{ref}}$, where $V_{\mathrm{ref}}$ is defined as the terminal sliding velocity for the case with $\mu_w = \mu_{\mathrm{ref}}$.

As shown in Fig. 7(a), for high-viscosity droplets, viscous dissipation occurs predominantly within the droplet. In this regime, the terminal velocity decreases with increasing droplet viscosity, following $V_t \propto \mu_w^{-1}$, and eventually approaches $V_t \approx V_{\mathrm{ref}}$. In both simulations [63] and experiment [41], this transition occurs at approximately $\mu_w \sim 2\mu_o$. Besides, we extend the case with $\mu_w = \mu_o$ in Fig. 7(b) to different $\mu_o$, $\theta_{ws}$, $\phi$, and $h_p$ to assess parameter sensitivity. The results show that variations in $\mu_o$ and $\theta_{ws}$ have a dominant effect on droplet translational mobility. By contrast, the geometric parameters have little influence, even when $\phi$ and $h_p$ are reduced by half. This observation is consistent with the previous numerical findings [63].

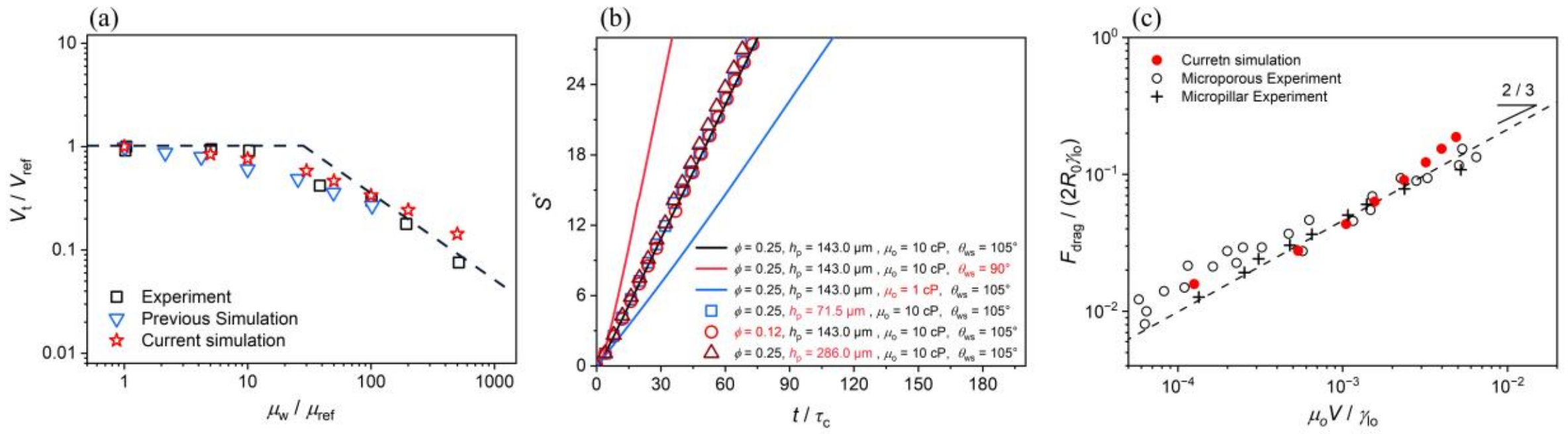

Figure 7. Simulation of droplet sliding on LIS driven by a constant body force: (a) comparison of our simulation results with experimental data [41] and previous numerical results [59] [63] for the droplet terminal velocity $V_t$; (b) parameter sensitivity analysis with respect to $\mu_o$, $\theta_{ws}$, $\phi$, and $h_p$. (c) Validation of the normalized drag force $F_{\text{drag}}/(2\gamma_{ow}R_0)$ as a function of $\mu_o V/\gamma_{lo}$ during droplet sliding on microporous [20] and micropillar LIS [42].

In addition to the droplet dissipation dominated regime as discussed above, we also performed simulations to validate $F_{\text{drag}}$ in the lubrication dissipation regime. In this case, similar to the experiment setup in Ref. [42]. we set $R_0 = 0.5\text{mm}$, $\mu_o = 20$ cP, $\phi = 0.5$, $\theta_{ws} = 105°$, and $\theta_{os} = 5°$. The droplet was driven by a constant horizontal force ranging from 2 to 10 μN, and was allowed to accelerate until reaching a steady state. Figure 7(c) presents the quantitative relationship between $F_{\text{drag}}$, and Ca. Our simulations successfully reproduce the $F_{\text{drag}} \sim \text{Ca}^{2/3}$ scaling law and agree well with the experimental results for microporous [20] and micropillar LIS surface [42]. The agreement with the experimental measurements provides targeted validation of the present numerical framework for the micropillar LIS considered in this study. Nevertheless, more stringent parameter-by-parameter validation over broad ranges of surface wettability, lubricant properties, and pillar geometry is currently limited by the lack of fully characterized experimental data. Targeted experiments with independently controlled and fully characterized parameters will therefore be pursued in future work.

# Data and code availability

The data and code generated in this study are available from the lead contact upon reasonable request.

# Acknowledgements

Support from State Key Laboratory of Space Medicine, China Astronaut Research and Training Center (Grant No. SLK2024K10), National Natural Science Foundation of China (No. 12502317), National Key R&D Program of China (Grant No. 2022YFF0503501), the UK Engineering and Physical Sciences Research Council under the project "UK Consortium on Mesoscale Engineering Sciences (UKCOMES)" (Grant No. EP/X035875/1) and "Multiphase Multicomponent Lattice Boltzmann Method for Modelling Wetting on Liquid Infused Surfaces" (Grant No. EP/V034154/2), Leverhulme Trust (Research Project RPG-2022-140) are gratefully acknowledged.